# Astro2020 Science White Paper
# Fundamental Physics with Brown Dwarfs: The Mass-Radius Relation

**Thematic Areas:**  ☐ Planetary Systems   ☐ Star and Planet Formation
☐ Formation and Evolution of Compact Objects   ☐ Cosmology and Fundamental Physics
☒ Stars and Stellar Evolution   ☐ Resolved Stellar Populations and their Environments
☐ Galaxy Evolution   ☐ Multi-Messenger Astronomy and Astrophysics

**Principal Author:**
Name: Adam Burgasser
Institution: UC San Diego
Email: aburgasser@ucsd.edu
Phone: +1 858 822 6958

**Co-authors:**
Isabelle Baraffe, University of Exeter
Matthew Browning, University of Exeter
Adam Burrows, Princeton University
Gilles Chabrier, University of Exeter
Michelle Creech-Eakman, New Mexico Tech
Brice Demory, University of Bern
Sergio Dieterich, Carnegie DTM
Jacqueline Faherty, American Museum of Natural History
Daniel Huber, University of Hawaii
Nicolas Lodieu, Instituto de Astrofisica de Canarias
Peter Plavchan, George Mason University
R. Michael Rich, UC Los Angeles
Didier Saumon, Los Alamos National Laboratory
Keivan Stassun, Vanderbilt University
Amaury Triaud, University of Birmingham
Gerard van Belle, Lowell Observatory
Valerie Van Grootel, University of Liege
Johanna M. Vos, American Museum of Natural History
Rakesh Yadav, Harvard University

**Abstract:**
The lowest-mass stars, brown dwarfs and giant exoplanets span a minimum in the mass-radius relationship that probes the fundamental physics of extreme states of matter, magnetism, and fusion. This White Paper outlines scientific opportunities and the necessary resources for modeling and measuring the mass-radius relationship in this regime.

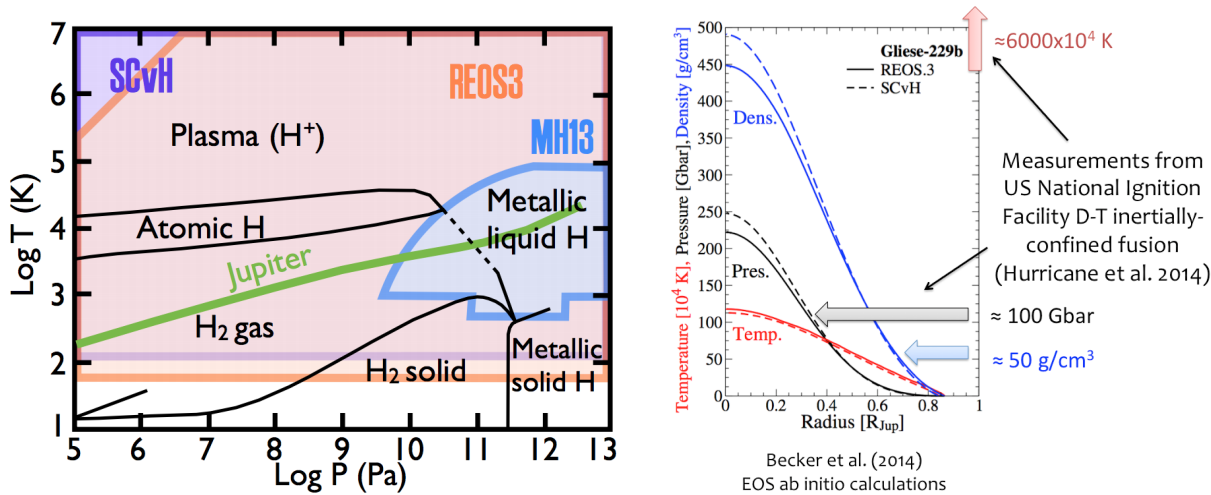

Figure 1: LEFT Phase diagram of hydrogen [35], indicating phase transitions (black lines) and ranges of validity for some published EOSs (SCvH = [48]; MH13 = [36]; RES03 = [5]). The internal pressure-temperature profile of Jupiter is indicated; brown dwarfs and very low mass stars have profiles that are progressively hotter and reach higher pressures (note: $10^5$ Pa = 1 bar). RIGHT: EOS ab initio model for Gliese 229B [5] compared to parameters probed by the US National Ignition Facility [25].

## Context

Brown dwarfs are star-like bodies unable to achieve the necessary core temperatures to sustain hydrogen fusion [29,40]. With core densities reaching 1000 g/cm$^3$ and core pressures reaching 1 Tbar ($10^{17}$ Pa), but temperatures below the proton-fusion threshold (T ≤ $10^6$ K [11]), these objects are supported by electron degeneracy pressure, and theory predicts that their interiors are composed of extreme states of matter, including metallic phases of hydrogen (Figure 1). While current inertially-confined fusion experiments can now achieve these pressures and densities, they do so at much higher temperatures (>>$10^6$ K; [25]). Thus, giant planets, brown dwarfs, and the lowest-mass stars provide an important opportunity to test theoretical models of extreme hydrogen equations of state (EOS) through mass, radius and density measurements.

The brown dwarf and giant exoplanet mass regime spans a minimum in the hydrogen-rich mass-radius relationship, from strongly degenerate (R ∝ M$^{-⅓}$) to mixed ideal/degenerate (R ∝ M) EOSs. However, the hydrogen EOS is not the only factor shaping this relation. Helium and metallicity contributions to the EOS and fusion efficiency; atmospheric opacity (including clouds) that modulates entropy loss; magnetic effects on pressure support, convective flow and surface spotting; and the age of an object all contribute to its particular radius. Burrows et al. [12] describe the mass-radius relationship as a "multi-parameter theory". Indeed, the few direct measures of radius in the brown dwarf mass regime obtained to date, while generally consistent with theoretical predictions, show considerable scatter (Figure 2), and non-EOS effects alone produce up to 10-25% variations in radius at a given mass and age in the models.

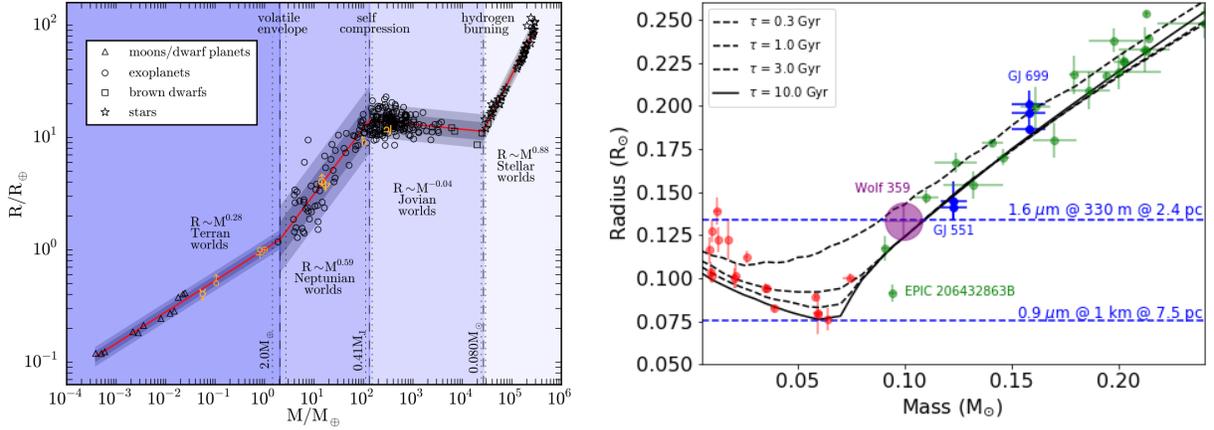

FIGURE 2: LEFT: The mass-radius relationship in context, from terrestrial planets to stars. Exoplanet transit measurements have provided radii from roughly Jovian masses downward, while eclipsing binaries and interferometry have provided radii in the stellar regime. At the high-density hydrogen fusion mass limit (M ≈ 0.072 $M_\odot$ ≈ 75 $M_{Jup}$ ≈ 2.4x10$^4$ $M_\oplus$), measurements are notably sparse [15].
RIGHT: Focusing on this region, transiting exoplanets (red) and eclipsing binaries (green) make up the bulk of current measurements, while interferometry(blue) just reaches the nearest very low mass star, Wolf 359 at 2.4 pc (purple; upper line shows limit for the CHARA array). A 1 km infrared array could potentially measure the entirety of the mass-radius relation (data from [4,7,58] and others).

**In the next decade, we will have the opportunity to refine measurements of the mass-radius relationship at the hydrogen fusion limit, and quantify the influence of various physical properties, enabling studies of extreme states of matter inaccessible in the laboratory.**

### Exploring the Mass-Radius Relation Through Current Observations

The majority of radius measurements in the substellar regime are for giant exoplanets or brown dwarfs on close, transiting orbits around more massive host stars (Figure 2). The structural evolution of these objects may be influenced by irradiation or tidal heating; or, if formed through core accretion, significantly supersolar metallicities [52]. Both unusually large and small systems have been identified; yet, despite model predictions, no robust empirical correlations have yet been found between radius deviation and mass, orbital period, irradiation, or metallicity in the low-mass stellar regime [51,62]. While transiting "low-mass" giant exoplanets (M ≤ 25 $M_{Jup}$) show clear correlations with irradiation [17,30], there is no such correlation at higher masses [4], and the few transit radius measurements near the hydrogen-burning limit inhibit robust statistical analysis of other trends. Current (e.g., KELT [43], *TESS* [45]) and future transit missions will significantly increase the number of transiting brown dwarfs, as long as sufficient resources are available for radial velocity follow-up to measure masses.

In the substellar regime, only one eclipsing brown dwarf system has been identified to date, 2MASS J0535-0546AB [54]. As a 1 Myr-old member of the Orion Nebula Cluster, the interiors of the brown dwarfs in this system have not yet reached degeneracy. Their radii also appear to be influenced by other effects, notably strong magnetism [44,55]. The faint luminosities of cooler, fully evolved brown dwarfs - spectral types L, T and Y - have been a persistent barrier to the detection of eclipsing systems in the field. Despite their abundance in the Galactic environment (>15% of all stars [28]), cool brown dwarfs are sparsely distributed in both shallow infrared synoptic surveys (e.g., *WISE*, PanSTARRS) and deep optical synoptic surveys (e.g., ZTF, DES). These eclipsing systems are particularly valuable as they provide both masses and radii. On the other hand, eclipsing systems among intermediate-aged systems (10-100 Myr) can also probe the evolution of the mass-radius with time, so cold brown dwarfs in the field and distant, warm brown dwarfs in young clusters both remain important targets.

One of the few significant correlations with radius variations in the low-mass regime is magnetic activity [32,55], which may explain the up to 10% inflation in radii observed among single and binary very low-mass stars relative to models [57]. This deviation may be due to magnetic pressure support [33,38], interior convective suppression [14,19], or dark magnetic spots [53]. As optical and X-ray signatures of magnetic emission - but not necessarily magnetic field strength - decline among cool brown dwarfs, correlating mass, radius and measures of magnetism (e.g., radio emission, Zeeman splitting) will be essential for clarifying the role that magnetic fields play on shaping the low-mass mass-radius relationship.

Indirect measurements of radii have also been critical. The Stefan-Boltzmann relation provides radii if a star's bolometric luminosity and effective temperature ($T_{eff}$) are known; adding surface gravity yields an object's mass [8]. Spectral surface flux model fits to absolute flux-calibrated spectra can also map objects on the mass-radius plane [31,50]. These methods are highly dependent on the fidelity of atmosphere modeling, and results have been mixed. Sorahana et al. [50] inferred radii as low as 0.64 $R_\odot$ from infrared model fits to L and T dwarfs, well below evolutionary models; Dieterich et al. [18] found a radius minimum among a well-characterized set of late-M and L dwarfs, but at $T_{eff}$s well above those predicted by evolutionary models. The persistent disagreements between structure models and observations based on these techniques likely require better calibration of atmosphere models and empirical relations.

## New Observational Opportunities

In addition to eclipsing pairs, monitoring observations of the lowest-mass stars and brown dwarfs can uncover transiting exoplanetary companions. The transit lightcurve geometry yields the average density of the host star [49], an alternative avenue for testing structure models. For example, the density of TRAPPIST-1 is lower than predicted by theoretical models, suggesting radius inflation comparable to other very low-mass stars (Figure 3, [10,21]). Again, this deviation may be due to magnetic [39] or metallicity effects [59], and it directly influences the inferred radii and densities of its exoplanet companions.

Another geometric approach to radius is the ratio of rotational velocity to variability frequency. While related through a typically unknown inclination angle, such measurements can be used to constrain lower radius limits [24], or statistically infer an underlying radius or radius distribution assuming a random inclination distribution [61]. Rotational velocities for resolved binaries can also provide radius constraints assuming spin-orbit alignment [23,28], although such alignment is not universally true among stellar binaries [1].

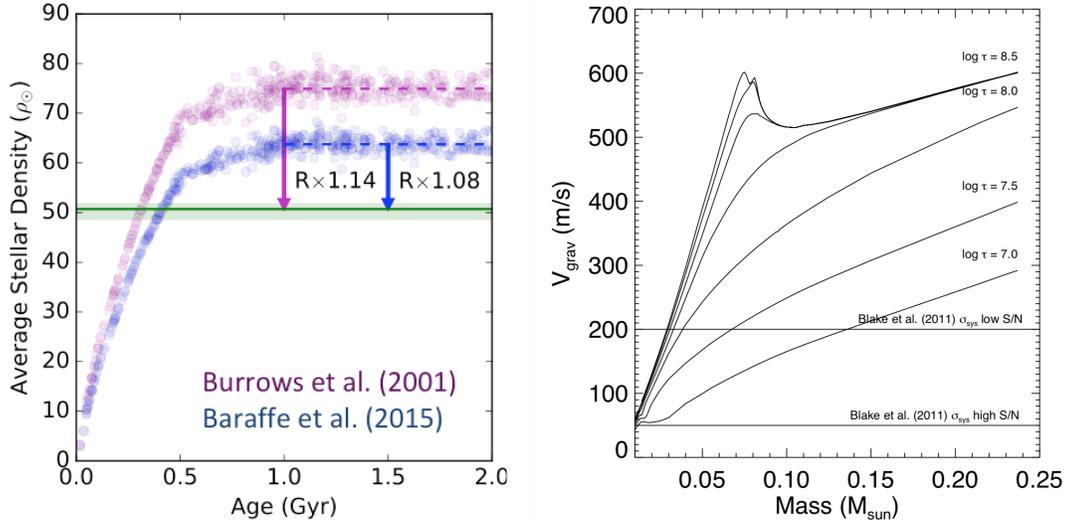

FIGURE 3: LEFT: Average stellar density of TRAPPIST-1 (green band [21]) compared to model predictions as a function of age at fixed $T_{eff}$. As the age of this system appears to be > 5 Gyr, evolutionary models underpredict the radius of this source by ≈10% (adapted from [10]). RIGHT: Gravitational redshift as a function of mass and age (log τ in years), based on models from [11]. At ages > 100 Myr, $V_{grav}$ varies minimally down to 0.07 $M_\odot$, then declines rapidly in the degenerate regime. RV precisions for M & L dwarfs using Keck/NIRSPEC [6] are shown.

An as-yet-to-be exploited opportunity in exploring the substellar mass-radius relation is the measurement of gravitational redshift, $V_{grav}$ = 0.64 (M/$M_\odot$)/(R/$R_\odot$) km/s [60]. Gravitational redshift varies minimally in the low-mass stellar regime, but declines sharply with mass below the hydrogen-fusion limit (Figure 3). Differential gravitational redshifts can be discerned in binaries with unequal mass components, a tactic previously exploited for white dwarfs in double degenerate systems [22] and in white dwarf-M dwarf pairs [42]. For a star-brown dwarf binary, gravitational redshift is potentially measurable in systems with precisely-constrained orbits ($V_{grav} \leq V_{orbit}$), or in widely-separated systems ($V_{grav} \gg V_{orbit}$), particularly with subgiant or giant star primaries. Mass-dependent gravitational redshift may be discernible in old open clusters, although dynamical scattering may mask the signal.

## Planning for the Next Decade

Improving our understanding of the mass-radius relationship from giant exoplanets to very low mass stars requires a significant increase in precision radius measurements for a variety of targets using multiple methods to fully explore physical trends, coupled with theoretical advances in the roles of magnetism, metallicity, clouds and the EOS on internal structure. We encourage the Decadal review committee to prioritize the following opportunities:

**Expand the sample of very low-mass eclipsing binaries and transiting exoplanet systems through deep-wide and/or targeted infrared synoptic surveys**: *Kepler* facilitated significant advances in stellar exoplanet demographics, asteroseismology, and eclipsing binaries, but the lowest-mass stars and brown dwarfs were simply too optically faint to be surveyed in sufficient numbers. *TESS* will be similarly limited, while ZTF and LSST cadences will rarely capture eclipse signatures. A deep infrared synoptic survey, perhaps modeled on the VISTA VVV [37], or extending targeted surveys such as MEarth [41] and TRAPPIST/SPECULOOS/Saint-Ex [16,20.46], would discover dozens of young and evolved brown dwarf eclipsing systems to populate the mass-radius relationship, uncover transiting exoplanets, and measure rotation periods.

**Pursue long-baseline infrared interferometry**: Current interferometry facilities are at the cusp of measuring the radii of nearby brown dwarfs. The 331m baseline CHARA array [56] has the sensitivity and resolution to resolve the 0.1 $M_\odot$ star Wolf 359 at 2.4 pc (Figure 2). Reaching brown dwarfs requires wider baselines and greater infrared sensitivity. An infrared system with 1 km baselines sensitive to J = 13-15 sources - feasible with today's technology - would resolve dozens of brown dwarfs out to 7-8 pc (and hundreds of stars), broadly populating the mass-radius plane and enabling evaluation of magnetic activity, cloud, and metallicity trends. These would also include binaries with precision *interferometric* orbits that can serve as benchmarks for testing evolutionary and atmosphere models (see Dupuy et al. WP).

**Enable High Resolution Spectroscopy for Faint Infrared Targets**: While radial velocity orbits have been measured for a few "hot" brown dwarf binaries with existing facilities [3,9,26,37,54], the rarity of eclipsing systems, and the fact that most fully-evolved brown dwarfs are low-temperature T and Y dwarfs, means that building the sample of eclipsing substellar binaries will require observations on large aperture (20-30m) telescopes such as TMT and GMT. Fortunately, the precisions required (≤ 0.5 km/s) are readily achievable, although higher precisions (< 10 m/s [13,34]) would enable mass measurements for exoplanet companions. High resolution spectroscopy are also needed for vsini and magnetic Zeeman broadening measurements.

**Advance interior structure theory through modeling of magnetic, metallicity, and atmospheric effects**: While empirical magnetic activity corrections for radii and $T_{eff}$ for M stars have proven valuable [55], a robust understanding of the role magnetic fields play in stellar structure requires continued theoretical development. This includes 3D magnetohydrodynamic models to account for both dynamo generation and structural/evolutionary effects in regions of strong degeneracy. Further exploration of metallicity effects on the interior EOS are needed and can be tested in compositionally-diverse systems, such as halo subdwarfs and open/globular cluster brown dwarfs. The entropy-controlling role of clouds and gas chemistry in low-temperature atmospheres requires further exploration and appropriate empirical test samples.